\documentclass{article}
\usepackage{graphicx} % Required for inserting images

\usepackage{multicol}
\usepackage[a4paper, margin=20mm]{geometry}
\usepackage{indentfirst}
\usepackage{authblk}

\usepackage{amsfonts}
\usepackage{amsmath}
\usepackage{float}
\usepackage{booktabs}
\usepackage{tabularx}
\usepackage{makecell}
\usepackage{longtable}
\usepackage{url}
\usepackage{abstract}
\usepackage{changepage}
\usepackage{enumitem}
\renewenvironment{abstract}{%
  \begin{adjustwidth}{-0.8cm}{-0.8cm} % 调整左右边距
  \noindent\textbf{} % 自定义标题
  \quotation
}{%
  \endquotation
  \end{adjustwidth}
}
\newcommand\keywords[1]{\textbf{Keywords}: #1}
\title{Dual Attention Residual U-Net for Accurate Brain Ultrasound Segmentation in IVH Detection}

\author[1]{Dan Yuan\thanks{E-mail: yuandan@cqepc.edu.cn}}

\author[2]{Yi Feng}
\author[3]{Ziyun Tang}

\affil[1]{Chongqing Electric Power College,Chongqing, China}
\affil[2]{Chongqing Metropolitan College of Science and Technology, Chongqing, China}
\affil[3]{Xinqiao Hospital, Army Medical University, Chongqing, China}
\date{}
\begin{document}
\maketitle
\begin{multicols}{2}
\begin{abstract}
\textbf{\textit{Abstract——}} 
Intraventricular hemorrhage (IVH) is a severe neurological complication among premature infants, necessitating early and accurate detection from brain ultrasound (US) images to improve clinical outcomes. Although recent deep learning methods offer promise for computerized diagnosis, challenges remain in capturing both local spatial details and global contextual dependencies critical for segmenting brain anatomies. In this work, we propose an enhanced Residual U-Net architecture incorporating two complementary attention mechanisms: the Convolutional Block Attention Module (CBAM) and a hybrid attention layers (HAL) made up of the Sparse Attention Layer (SAL) and Dense Attention Layer (DAL). The CBAM improves the model's ability to refine spatial and channel-wise features, while the HAL introduces a dual-branch design, sparse attention filters out low-confidence query-key pairs to suppress noise, and dense attention ensures comprehensive information propagation. Extensive experiments on the Brain US dataset demonstrate that our method achieves state-of-the-art segmentation performance, with a Dice score of 89.04\% and IoU of 81.84\% for ventricle region segmentation. These results highlight the effectiveness of integrating spatial refinement and attention sparsity for robust brain anatomy detection. Code is available at: https://github.com/DanYuan001/BrainImgSegment.
\end{abstract}
\section{Introduction}
According to the World Health Organization, 15 million babies are born prematurely each year {}\cite{2013Born,2023RoofBag,2024Landslide}. For premature infants, one of the most common brain injuries is intraventricular hemorrhage (IVH) {}\cite{2012Neonatal}. These hemorrhages result in ventricle dilation, which can cause serious brain damage if not properly treated. Therefore, we must continue to monitor the ventricle dilation to detect the occurrence of IVH as early as possible. From the point of view of machine vision technology, the task is to accurately segment the ventricle from an ultrasound image to obtain the anatomical structures of the brain.

Due to advances in deep learning networks, medical image analysis has undergone a tremendous revolution. There are many systems that conduct the anatomy segmentation process in ultrasound images, assisting radiologists in an accurate diagnosis. For instance, the breakthrough of convolutional neural networks (CNNs) in image classification and recognition has increased for automatic segmentation and recognition of anatomy in ultrasound images. The features of the various image data can be extracted and segmentation can be performed in ultrasound images {}\cite{2021Medical}. In addition, the U-Net series emerged in a variety of case studies using computer vision approaches in the medical imaging field, such as U-Net {}\cite{2015U}, U-Net++ {}\cite{2018UNet}, V-Net {}\cite{2016V} U-Net3+ {}\cite{2020UNet}, Res-UNet {}\cite{2018Weighted} and KIU-Net{}\cite{2020KiU}. These variant models of U-Net consist of a basic encoder-decoder structure, with the encoder extracting features from the input images and the decoder producing a segmentation mask through upsampling in each level. The U-Net series have shown excellent performance in various public and private medical datasets. However, CNNs lack the ability to capture the correlation of global image features due to the inherent spatial local inductive bias. More precisely, each convolutional kernel only processes a local subset of pixels in the entire image and makes the network concentrate on local patterns rather than global context.

In various computer vision tasks, attention mechanisms have produced promising results. The attention framework {}\cite{2020Axial} and {}\cite{2021Medical}, for instance, introduces attention gates to emphasize relevant features and suppress irrelevant ones. Other attention mechanisms, such as sparse attention {}\cite{zhang2021sparse}, have been utilized to highlight positive correlations between image tokens and filter out negative ones with the cooperation of the dense layer and the sparse layer. Some more attention mechanisms, such as the channel attention (CA) and spatial attention (SA) mechanism {}\cite{2020Spatial} have also been introduced to better describe the features extracted from images in learning networks. As we have discussed that CNNs are only capable of describing local context and so is the U-Net series, capturing global features of images has been a challenge in order to improve the vision system performance. Studies such as multilevel pyramid structure {}\cite{2024FCPFNet}, multiscale approach like {}\cite{2022AA}, and {}\cite{2023PCDM} have introduced various image "zoom-in" models to achieve the goal.

However, the key problem of the full attention mechanism is that it computes not only in some informative regions , but most of the uninformative ones of an image or feature map. The modified sparse attention mechanism in this paper has the ability to filter out the negative impact between queries and keys, and it is still able to capture the long-range data correlation, that is, its receptive field range is not weaker than that of the full attention models. 

\begin{figure*}
    \centering
    \includegraphics[width=\linewidth]{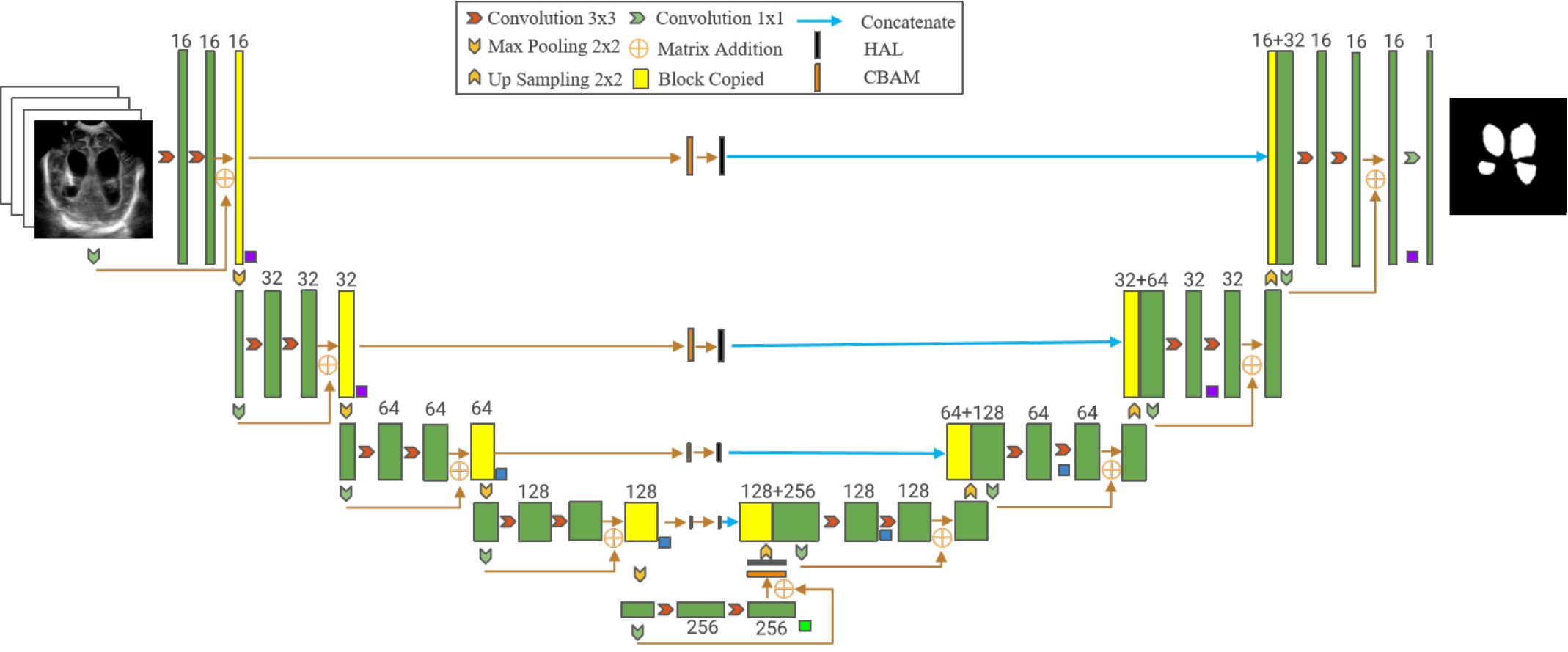}
    \caption{\textbf{Block diagram of the proposed model used for medical image segmentation. An input image with dimensions 128 × 128 × 1 undergoes feature extraction through the encoder, and the decoder then performs upsampling on the encoded features to predict a binary mask of size 128 × 128 × 1. The in-between connections of the encoder and the decoder are accompanied by the addition of CBAM and HAL to enhance the performance.}}
    \label{fig:Fig1}
\end{figure*}

In this paper, we propose a novel segmentation method for detecting brain anatomies in brain ultrasound images. The approach is a Residual U-Net framework consisting of two attention mechanisms, namely the convolutional block attention module (CBAM) {}\cite{2018CBAM} and the hybrid attention layer (HAL), which not only better describe channel-wise features and spatial-wise relationships, but also neglect the correlations from uninformative regions and highlight informative ones. The general framework of the proposed segmentation model is shown in Fig. \ref{fig:Fig1}.

The contributions of this work are as follows.
\begin{itemize}
\item We utilize a revised hybrid attention layer (HAL), which selectively manipulate the correlations between data tokens, and is integrated into the proposed segmentation model;
\item We integrate into the model the CBAM attention mechanism which captures both the channel-wise features and spatial-wise relationships, making the feature representations more accurate and interpretable.
\item The series connection of CBAM and HAL in the entire framework significantly improves the performance of the model on both the Brain US dataset and a supplementary dataset (Nuclei). The entire model achieves state-of-the-art performance. 
\end{itemize}

\section{Related Work}
Intraventricular hemorrhage (IVH) remains a critical global health concern, particularly affecting premature infants. To improve early alarming and diagnosis, numerous deep learning-based approaches have been explored in recent years. These methods have shown substantial progress in segmentation and classification tasks in a wide range of medical imaging modalities~\cite{2020Learning, 2020KiU, 2022AAU, 2021Medical}. For instance, the Confidence-Guided Brain Anatomy Segmentation Network (CBAS)~\cite{2020Learning} estimates both segmentation masks and corresponding confidence maps on multiple scales, and leverages a confidence-weighted learning mechanism to enhance model reliability. Similarly, the work in~\cite{2020KiU} proposes an over-complete convolutional architecture to project input images into higher dimensions, effectively controlling receptive field growth and enhancing small-structure detection.

Recent efforts~\cite{2024IMAGPose, 2025IMAGGarment} have begun to explore modular attention frameworks tailored to domain-specific constraints. IMAGDressing-v1~\cite{2025IMAGDressing} introduces controllable visual pipelines integrating multilevel attention to achieve generation of structural consistency. Similarly, motion-aware modeling frameworks, such as MCDM~\cite{2025LongTalkingFace} emphasize the importance of combining spatial structure, motion priors, and context-level awareness. Based on these insights, our work investigates the integration of channel-wise, spatial, and class-aware attention into a unified segmentation architecture optimized for IVH-related brain ultrasound imagery.

In parallel, hybrid attention mechanisms have gained popularity for improving model focus and generalization. AAU-Net~\cite{2022AAU} integrates adaptive convolutions with channel and spatial self-attention to replace standard convolution operations, showing improvements in breast ultrasound segmentation. The Convolutional Block Attention Module (CBAM)~\cite{2018CBAM} has also proven effective in guiding networks to attend to discriminative features, and has been widely adopted in both general vision and medical imaging domains. More recently, positional and multiscale attention mechanisms have been used to further model spatial dependencies~\cite{2022Multiscale}, while transformer-based networks~\cite{2021Medical} have incorporated gated axial attention into residual U-Net backbones for improved context modeling. Despite these advancements, many existing architectures still struggle with balancing local spatial cues and global semantic representations, especially in ultrasound-based anatomical segmentation.

\section{Methodology}
The research shows an overall framework that integrates both CBAM and HAL based on the residual U-Net (Res-UNet) scheme. The Res-UNet consists of the encoder and decoder, which work in a U-shaped pipeline (in Fig. \ref{fig:Fig1}), i.e. extract features from the original image and generate segmentation results.
\subsection{Encoder of Residual U-Net}

The encoder has four layers with multiple 3×3 convolution kernels and a stride of 1, which are common parameter configurations in U-Net. In each layer, convolutions are followed by batch normalization and the ReLU activation function for the network. After that, skip connections are introduced for the compensation to avoid the problem of gradient vanishing. In most of the U-Net framework, a ratio of 2 is applied for downsampling across the levels. In addition, the serialized attention modules HAL and CBAM, are fed with features output from the encoders to ensure the feature enhancements before connecting to the decoders via skip connections. Such a feature-intercepting scheme has been widely used in many case studies using U-Net frameworks, providing a way to refine the features before reaching the decoders.
\subsection{Decoder of Residual U-Net}

The decoder acts as the reverse process of the encoder. It also needs to combine upsampled feature maps with those from the attention layers at each level, including both low-level and high-level features. In addition, the batch normalization and activation functions are indispensable, with the same configuration as the encoder. Moreover, the HAL layer is incorporated in series with the CBAM, capturing global dependencies and filtering out the uninformative features in the process of attention computation.
\subsection{Feature Fusion and Attention}

\begin{figure*}
    \centering
    \includegraphics[width=\linewidth]{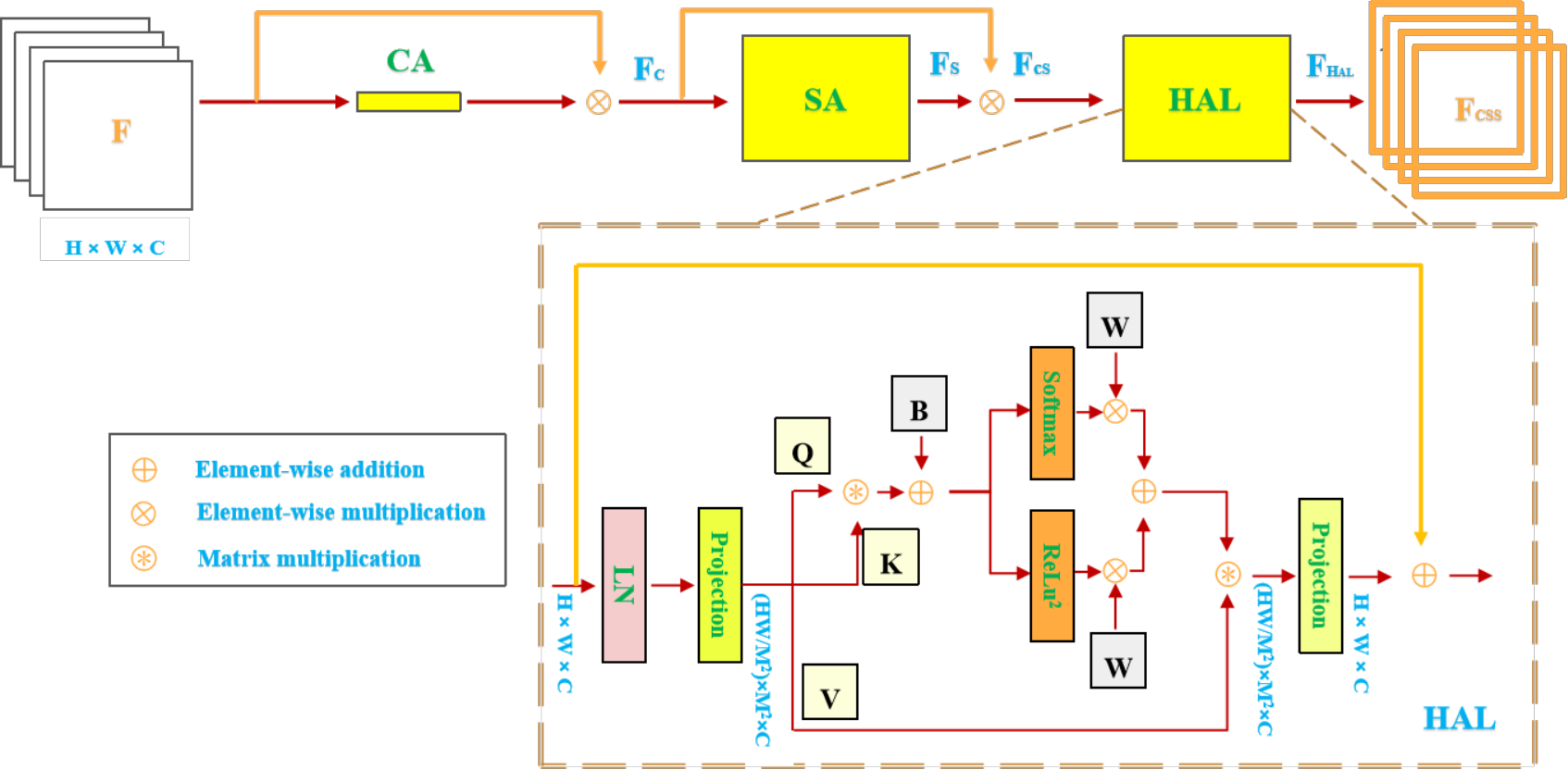}
    \caption{An illustration of CBAM and HAL blocks. CBAM and HAL are applied to the input feature F. The addition of the outputs of CBAM and HAL is fed to the decoder. Note that the module "HAL" is composed of a sparse layer and a dense layer.}
    \label{fig:Fig2}
\end{figure*}

Following feature extraction via the encoder, we perform adaptive feature enhancement using attention mechanisms. The CBAM attention module comprises a sequence of efficient channel attention (CA) and spatial attention (SA) components, as well as HAL. The network details are shown in Fig. \ref{fig:Fig2}.
\subsubsection{Channel and Spatial Attention}

Given a feature map $F\in\mathbb{R}^{H\times W\times C}$ , where \textit{H}, \textit{W}, and \textit{C} is the height, width, and number of channels, respectively. CA assigns weights to the channels on the feature map, enhancing specific channels that more contribute to improving model performance. \textbf{F} is successively processed by a one-dimensional channel attention module and a two-dimensional spatial attention module, this process is similar to {}\cite{2018CBAM} and {}\cite{2024DAUNet}, and can be formulated as follows
\begin{equation}
\begin{split}
F_C=\sigma(MLP(AP(F))+MLP(MP(F)))\otimes F ,\\
F_S=\sigma(K^{7\times 7}[AP(F_C);MP(F_C)]) ,\\
F_{CS}=F_S\otimes F_C .
\end{split}
\label{Eq:Eq1}
\end{equation}
where $\sigma$ represents sigmoid activation function, AP denotes average pooling, MP denotes max pooling and $\otimes$ denotes element-wise multiplication. $K^{7\times 7}$ is a large kernel with a dilation of 4. $F_C\in\mathbb{R}^{H\times W\times C}$ and $F_S\in\mathbb{R}^{H\times W\times C}$ represent the features after the channel and spatial selections, and $F_{CS}\in\mathbb{R}^{H\times W\times C}$ is the output after these two modules. The CA module captures channel-wise dependencies and SA is for the spatial ones.

\subsubsection{Hybrid Attention Layer}
As the Transformers compute all tokens inside the feature map, it may compute in many irrelevant areas. Thus, it introduces irrelevant features, which reduces the accuracy and computational efficiency of the model. In order to deal with this issue, a type of sparse attention {}\cite{zhang2021sparse} layer is proposed to filter out the features with negative impacts on the map, i.e., the low relevance between queries and keys. Among all activation functions, the ReLU function is considered because it clears to zero in the negative range. In the meantime, there should be a dense attention layer (DAL) as a compensator if an oversparsity of attention occurs{}\cite{Shen2023study}. The DAL retains crucial information about the features. Unlike the sparse layer, the dense layer uses the softmax activation function. The idea of using this dual-branch scheme is to reduce the noisy information while retaining the informative one as much as possible. Therefore, the entire network copes with the fusion by using the features of these two branches and feeds them forward into successive modules.

Similar to the VIT {}\cite{2021An} partitioning process, the non-overlapping image patches are generated with a fixed size of $M\times M$, resulting in a flattened representation $F^i\in\mathbb{R}^{M^2\times C}$, where $F^i$ means the \textit{i}-th patch feature. Then the matrices of query \textit{\textbf{Q}}, key \textit{\textbf{K}}, and value \textit{\textbf{V}} are generated from \textit{\textbf{F}} by multiplying it with linear projection matrices, respectively. Linear projection matrices $W_Q$, $W_K$, and $W_V\in\mathbb{R}^{C\times d}$ are learnable and shared among all patches. The attention can be computed as follows
\begin{equation}
A=Activation(\frac{QK^T}{\sqrt{d_K}}+B)V.
\label{Eq:Eq2}
\end{equation}
where \textit{\textbf{A}} denotes the estimated attention result matrix, and \textit{\textbf{B}} is the learnable relative position bias. 

The regular dense attention layer (DAL) is used in most of networks by substituting the activation by softmax as follows
\begin{equation}
Att_1=softmax(\frac{QK^T}{\sqrt{d_K}}+B).
\label{Eq:Eq3}
\end{equation}

Since any query token is not likely to be closely relevant to all keys, calculating all similarities is not necessary for image segmentation. Therefore, designing a sparse attention layer to build useful correlations among tokens could improve feature fusion. Replacement of activation with a ReLU-based layer appears to be a plausible way to obtain the sparsity of attention {}\cite{zhang2021sparse} {}\cite{2024AP}. It feeds forward the useful information after filtering out the similarities with negative relevance:
\begin{equation}
Att_2=ReLU^2(\frac{QK^T}{\sqrt{d_K}}+B).
\label{Eq:Eq4}
\end{equation}

Note that a certain compensation technique is often demanded to overcome the sparsity. The feature fusion by building dual branches (SAL and DAL) is necessary.

SAL, as an irrelevance filter, may cause insufficient information problems for the following process. In contrast, DAL inevitably introduces redundant information in many irrelevant regions, posing an interference in image segmentation. Therefore, we prefer a dual branch hybrid self-attention mechanism.

Thus, by fusing both of the branches, the further weighted attention score can be defined as follows
\begin{equation}
F_{HAL}=(\sum_{i=1}^{2}(\omega_iAtt_i))V.
\label{Eq:Eq5}
\end{equation}
where $\omega_i\in\mathbb{R},i=1,2$ are two normalized weights for modulating two branches.

Finally, the attention layer takes the output from the CBAM, and its resulting attention maps are concatenated to the decoders.
\subsection{Loss Function Combinations}
There are quite a few loss functions suitable for image segmentation tasks. Dice loss, binary cross-entropy (BCE) loss, and focal loss are popular for our specific tasks. Although these loss functions are common in most textbooks, we still study the loss function selection strategy in the work {}\cite{2024DAUNet}, where various combinations of these three loss functions are tried. These loss functions monitor the training process by quantifying the similarity between the ground truth and the predicted masks.

The Dice loss, related to \textit{\textbf{F1}} score, is derived from the Dice coefficient, which calculates the intersecting area between the ground truth and the predicted masks. The Dice loss is formulated using \eqref{Eq:Eq6}.
\begin{equation}
l_{Dice}=\frac{FP+FN}{2\times TP+FP+FN}.
\label{Eq:Eq6}
\end{equation}
In (\ref{Eq:Eq6}), \textit{\textbf{TP}} stands for the number of true positive pixels, \textit{\textbf{FP}} for the number of false positive pixels, and \textit{\textbf{FN}} for the number of false negative pixels. 

The BCE loss minimizes the dissimilarity between the predicted and ground truth masks. The BCE loss can be formulated as  \eqref{Eq:Eq7}.
\begin{equation}
\begin{split}
l_{BCE}(I,\hat{I})=-\frac{1}{WH}(\sum_{i=0}^{W-1}\sum_{j=0}^{H-1}[I(i,j)log(\hat{I}(i,j))\\+(1-I(i,j))log(1-\hat{I}(i,j))]).
\end{split}
\label{Eq:Eq7}
\end{equation}

where $\textbf{I}$ represents the ground truth label , $\hat{\textbf{I}}$ represents the foreground prediction.

Focal loss is used to concentrate on difficult cases in image segmentation tasks. The larger weights are given to the hard part of the image pixels, while the smaller ones are assigned to the easy pixels during training. The Focal loss is formulated as follows
\begin{equation}
l_{Focal}=-\frac{1}{N}(\sum_{i=1}^{N}\alpha (1-p_i)^\gamma log(p_i)).
\label{Eq:Eq8}
\end{equation}

In (\ref{Eq:Eq8}),  $\alpha$ is a balancing parameter and $\gamma$ is the focusing parameter.

In training, we tried using a single loss function and also combinations of these functions, as shown in forms of \eqref{Eq:Eq9}  to monitor the optimization process. 
\begin{equation}
l=\alpha l_{Dice}+\beta l_{BCE}+\lambda l_{Focal}.
\label{Eq:Eq9}
\end{equation}
\section{Experiments and Results}
\subsection{Dataset Description}
We use the brain anatomy segmentation data set (Brain US) {}\cite{2020Learning}, {}\cite{2018Automatic} to evaluate our model. Intraventricular hemorrhage (IVH) resulting in enlargement of the brain ventricles is one of the main causes of preterm brain injury. The main imaging modality used for the diagnosis of brain disorders in preterm newborns is cranial ultrasound because of its safety and cost-effectiveness. In addition, the absence of septum pellucidum is an important biomarker for the diagnosis of septo-optic dysplasia. Automatic segmentation of the brain ventricles and septum pellucidum from these US scans is essential for an accurate diagnosis and prognosis of these diseases. After obtaining institutional review board (IRB) approval, US scans were taken from 20 different premature neonates (age less than 1 year). The total number of images collected was 1629 with annotations out of which 1300 were allocated for training and 329 for testing. We resize the images to 128×128 for all our experiments.
\subsection{Evaluation Metrics}
Tow metrics are used to evaluate the performance of our segmentation model: dice score (F1) and intersection over union score (IoU). These metrics provide a quantitative way to evaluate the performance of the segmentation process.

\subsubsection{Intersection over Union}  IoU is a measure that computes the overlap between the predicted segmentation mask and the ground truth mask. It is formulated as follows
\begin{equation}
IoU=\frac{TP}{TP+FP+FN}.
\label{Eq:Eq10}
\end{equation}
\subsubsection{Dice Score}  The Dice score, also referred to as the \textit{\textbf{F1}} score, is formulated as follows
\begin{equation}
F_1=\frac{2\times TP}{2\times TP+FP+FN}.
\label{Eq:Eq11}
\end{equation}

\subsection{Experimental Setup}
We have developed our segmentation model using Python and have used the PyTorch library for implementation. To speed up training, we used the high performance NVIDIA RTX 4060Ti 16G.

\subsection{Hyperparameter Setup}
Our model is trained for 150 epochs for the Brain US dataset. For the image size, we resize all images to a uniform size of 128×128 pixels, which are fed into the segmentation model. During the training phase, we use the Adam optimizer with a learning rate of 0.00001 to optimize the model. To ensure a comprehensive evaluation, we use the common split strategy and divide the data into a split: 70\% (train), 10\% (test) and 20\% (validation). All the setups discussed above are usually common in such a training task.

\begin{figure}[H]
    \centering
    \includegraphics[width=\columnwidth]{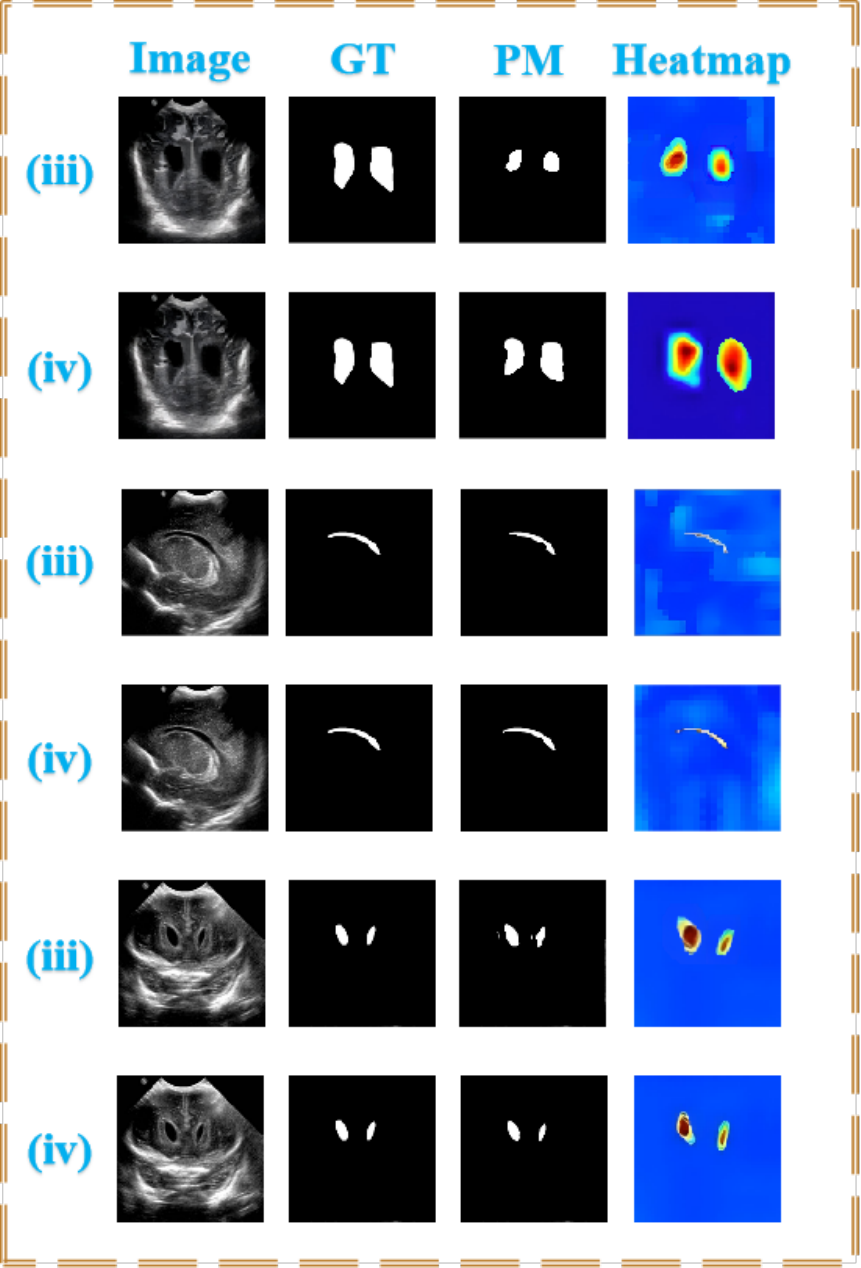}
    \caption{\textbf{Results of the ablation study indicate the improvement in model performance with each experimental modification.} GT and PM are ground truth and predicted mask, respectively. Heatmaps demonstrate the improvement of the model's performance in focusing on the region of interest.The SAL branch filters out the negative impact of query-key matching scores so that the heatmap looks monotonic in the uninformative regions. Note that there is no attention map for case i and case ii.}
    \label{fig:Fig3}
\end{figure}

\subsection{Ablation Tests}
\subsubsection{Upgrading Model from Scratch}  A series of experiments are conducted to refine our segmentation model and evaluate the impact of upgrading modifications. These experiments include:
\begin{enumerate}[label={\roman*.}]
\item Baseline U-Net model as the initial benchmark.
\item Residual U-Net model with skip connections.
\item Residual U-Net model with CBAM applied to the skip connections.
\item Proposed Res-UNet model with the strengths of CBAM and HAL.
\end{enumerate}
The results in Table \ref{tab:Tbl2} show the efficacy of each modification in the upgrading way. Each model has been trained using the linear combination of Dice, BCE, and Focal loss with average weights. Both CBAM and HAL increase Dice and IoU by around 3 points with respect to the baseline. Furthermore, compared to the pure Res-UNet case, the dual attention modules between encoders and decoders contribute about 1.5 to 2 point improvement in terms of both metrics.

Fig. \ref{fig:Fig3} visually illustrates the performance achieved with our combinations of CBAM and HAL attention mechanisms. The entire framework can refine feature representations and capture both global and local spatial correlations for accurate segmentation. Compared with the CBAM-only case (iii), the full model (iv) performs the segmentation more precisely because the SAL branch filters out the 'disturbing' information from the attention map. Furthermore, the heatmap visualization of the proposed model illustrates the spatial regions where it places more emphasis, showing a close resemblance to the ground truth regions for the Brain US dataset.

\subsubsection{Combinations of Loss Functions}  We have performed some incremental tests on different combinations of loss functions and have consolidated the results in Table \ref{tab:Tbl1}. Based on the results in Table \ref{tab:Tbl1}, we have compared the Dice and IoU scores for each combination. Our conclusion is that the linear combination of all three loss functions produces the best result in our segmentation tests. Therefore, we applied this paradigm for all the experiments in the paper.

\begin{figure*}
    \centering
    \includegraphics[width=\linewidth]{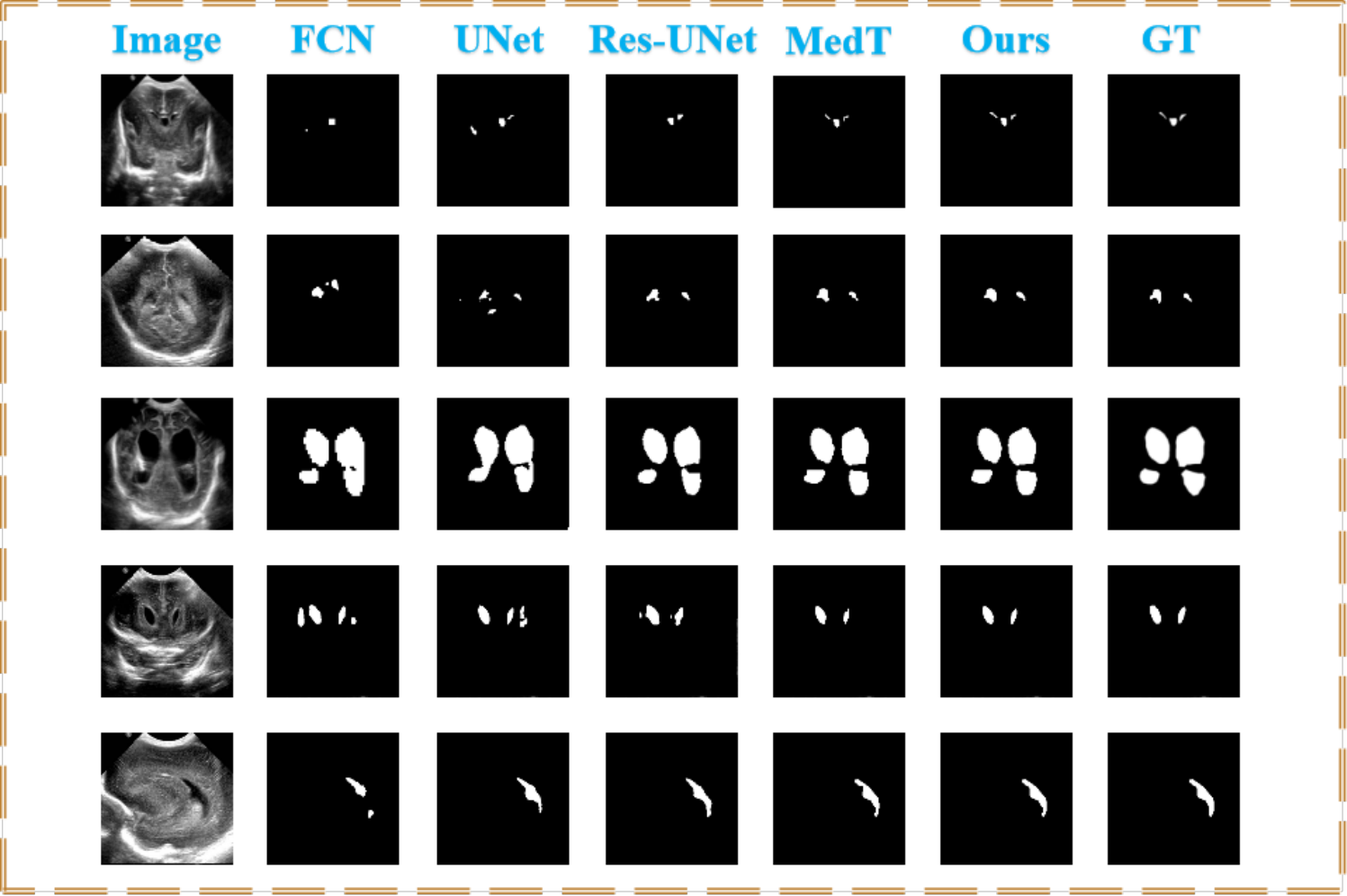}
    \caption{\textbf{Results of the proposed segmentation model on images of the Brain dataset.} All the results are obtained by running their respective code.}
    \label{fig:Fig4}
\end{figure*}

\subsection{State-of-the-art Comparisons}
We have performed a comparison between our proposed method and several baseline models along with standard segmentation models. The comparative results, which comprehensively evaluate two metrics, are presented in Table \ref{tab:Tbl3}. The models compared in the table are well known in the field of image segmentation, such as FCN {}\cite{2017SegNet}, U-Net {}\cite{2015U}, UNet++ {}\cite{2018UNet}, Res-UNet {}\cite{2018Weighted}, Axial Attention U-Net {}\cite{2020Axial}, and Medical Transformer {}\cite{2021Medical}. Our proposed method demonstrates superior performance across these models (see Table \ref{tab:Tbl3}) in terms of Dice score and IoU, indicating better overall segmentation accuracy. To provide specific performance details, our proposed model achieves a Dice score of 89.04, indicating a higher level of similarity between the ground truth and the predicted segmentation masks. In addition, the IoU metric, with a value of 81.84, indicates the strong capability of the model to accurately delineate regions of interest. In conclusion, the proposed model achieves the highest Dice scores out of all the models, suggesting that it performs better than other models.
\begin{figure}[H]
    \centering
    \includegraphics[width=\columnwidth]{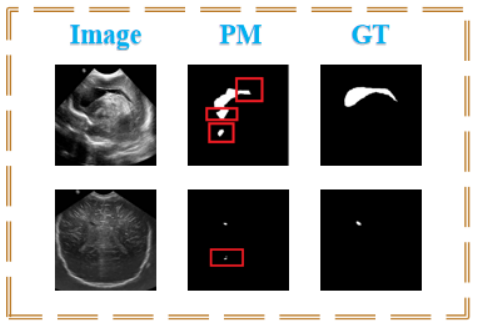}
    \caption{\textbf{Illustration of some of the failure cases of our model.} The enclosed regions are the misclassified segmented masks. GT and PM represent the Ground Truth and Predicted Mask, respectively.}
    \label{fig:Fig5}
\end{figure}

\begin{table*}[!t]
\setlength{\tabcolsep}{3mm}
\caption{\textbf{Performance metrics of the proposed model with different loss functions.} }
\label{tab:Tbl1}
\centering
\resizebox{\columnwidth}{!}{
            \begin{tabular}{ccc} 
                \toprule
			Loss Functions & Dice & IoU  \\
			\midrule
			BCE Loss & 70.73 & 74.91  \\
			%\midrule
			Dice Loss & 72.94 & 66.21  \\
                %\midrule
			Focal Loss & 67.16 & 61.40  \\
                %\midrule
			BCE Loss + Dice Loss & 87.32 & 78.89  \\
                %\midrule
			BCE Loss + Focal Loss & 74.61 & 76.27  \\
                %\midrule
			Dice Loss + Focal Loss & 83.24 & 75.02  \\
                %\midrule
			BCE Loss + Dice Loss + Focal Loss & \textbf{89.04} & \textbf{81.84}  \\
			\bottomrule
		\end{tabular}   
        }
\end{table*}

\begin{table*}[!t]
\setlength{\tabcolsep}{3mm}
\caption{\textbf{Dice scores of the segmentation models.} All values are in \%. Bold values indicate superior performance. The results are in x(+y) format, where x is the score and y is the improvement over the baseline.}
\label{tab:Tbl2}
\centering
            \begin{tabular}{ccccc} %设置表格每列的对齐方式,c表示居中,r表示右对齐,l表示左对齐
                \toprule
			Metrics & (i)U-Net[17] & (ii)Res-UNet[27] & \makecell{(iii)Res-UNet\\(w/CBAM only)} & \makecell{ (iv)Res-UNet\\(w/CBAM and HAL(Ours))} \\
			\midrule
			Dice & 85.37 & 87.50(+2.13) & 87.97(+2.60) & \textbf{89.04}(+3.67) \\
			%\midrule
			IoU & 79.31 & 79.61(+0.30) & 80.64(+1.33) & \textbf{81.84}(+2.53) \\
			\bottomrule
		\end{tabular}        
\end{table*}

\begin{table*}[!t]
\setlength{\tabcolsep}{1.3mm}
\caption{\textbf{Performance comparison with standard segmentation models.} All values are in \%. Bold values indicate superior performance.}
\label{tab:Tbl3}
\centering
            \begin{tabular}{cccccccc} 
                \toprule
			Metrics & FCN{}\cite{2017SegNet} & U-Net{}\cite{2015U} & U-Net++{}\cite{2018UNet} & Res-UNet{}\cite{2018Weighted} & \makecell{Axial Attention \\ U-Net{}\cite{2020Axial}} & \makecell{Medical \\Transformer{}\cite{2021Medical}} & \makecell{Proposed\\(Ours)} \\
			\midrule
			Dice & 82.79 & 85.37 & 86.59 & 87.50 & 87.92 & 88.84 & \textbf{89.04} \\
			%\midrule
			IoU & 75.02 & 79.31 & 79.95 & 79.61 & 80.14 & 81.34 & \textbf{81.84} \\
			\bottomrule
		\end{tabular}        
\end{table*}

Fig. \ref{fig:Fig4} shows the segmentation results of our proposed model, demonstrating its ability to segment the regions of interest in ultrasound images accurately. Although the visual result differences between our model and MedT are somehow tiny in most of the masks, it still achieves the best segmentation results by far. This suggests that the model pays attention to only relevant areas, which is beneficial to its segmentation accuracy.

By these comprehensive evaluations and visualizations, our proposed model demonstrates its ability to significantly improve the detection and diagnosis of brain anatomy , leading to the goal of early alarming.

\subsection{Weakness Analysis}
Our model has shown superior performance in various image segmentation tasks, as depicted in Table \ref{tab:Tbl3} and Fig. \ref{fig:Fig4}. However, it is essential to highlight the failure cases in which non-anatomical regions are misclassified as anatomical ones and vice versa, indicating that the precision and recall are relatively lower. Intuitively, the quality of segmentation results is dependent on the complexity of the dataset for training and evaluation. Fig. \ref{fig:Fig5} illustrates the difficulties encountered by our model, leading to deviations from the ground truth. These challenges may arise from the complexity of the dataset, variations in image quality, or the presence of ambiguous features that are difficult to delineate accurately. Despite these challenges, our proposed model has shown valuable contributions in the field of anatomical segmentation.
\begin{figure}[H]
    \centering
    \includegraphics[width=\columnwidth]{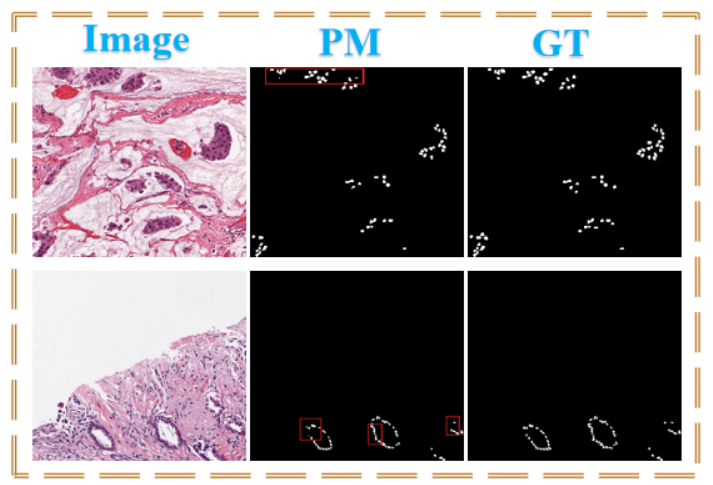}
    \caption{\textbf{Predicted mask visualization of the proposed model on the Nuclei dataset.} GT and PM represent the Ground Truth and Predicted Mask, respectively. The enclosed regions are the misclassified segmented masks.}
    \label{fig:Fig6}
\end{figure}

\subsection{Experimentation on the Nuclei Dataset}
In order to evaluate the effectiveness of our proposed method, we performed an evaluation on another dataset named Nuclei, which is a well-known collection that contains bright microscopy images of stained cells and correctly annotated masks by expert pathologists. The average resolution of the Nuclei images and their corresponding correct annotation masks is 2000×2000 pixels. For the evaluation of this dataset, we used the same set of hyperparameters as used in the evaluation of the Brain US data set to maintain consistency. The images should be converted to grayscale and resized to the same size as the images in the Brain US dataset before being fed into the network. The segmentation results of the proposed model on the Nuclei dataset are shown in Fig. \ref{fig:Fig6}. The visual results show that the overall performance is worse than the performance of the Brain US dataset, as the segmentation area is made up of many tiny grainy regions, which makes it difficult to focus attention.
\section{Conclusion and Future Research}
The current research proposes an innovative segmentation technique, called attention-based Res-UNet model, to detect brain anatomy in ultrasound images. Our model has provided satisfactory results, revealing its potential to improve detection and diagnosis. In the weaknesses section, some potential areas for future research were identified. One particularly promising way involves studying more advanced architectures that integrate multiple more powerful attention within the model. This approach has the potential to improve performance and expand our understanding of the complex challenges in detecting brain anatomy. Another direction for future research involves the dataset using techniques such as data augmentation, which can increase their diversity and size, leading to improved generalization and robustness. Furthermore, generalizing segmentation on other types of medical images, as we did for the nuclei data set, will assess the versatility and applicability of the model to a variety of medical imaging tasks.
\section{Acknowledgments}
This project is supported by the Scientific and Technological Research Program of ChongQing Municipal Education Commission (Grant NO. \textbf{KJQN202402609}).

\end{multicols}

\end{document}